# RCR: Robust Compound Regression for Robust Estimation of Errors-in-Variables Model

## Hao Han and Wei Zhu

Department of Applied Mathematics and Statistics, State University of New York, Stony Brook, NY 11794, USA

*Corresponding Email*: hanhao224@gmail.com

## Abstract

The errors-in-variables (EIV) regression model, being more realistic by accounting for measurement errors in both the dependent and the independent variables, is widely adopted in applied sciences. The traditional EIV model estimators, however, can be highly biased by outliers and other departures from the underlying assumptions. In this paper, we develop a novel nonparametric regression approach - the robust compound regression (RCR) analysis method for the robust estimation of EIV models. We first introduce a robust and efficient estimator called least sine squares (LSS). Taking full advantage of both the new LSS method and the compound regression analysis method developed in our own group, we subsequently propose the RCR approach as a generalization of those two, which provides a robust counterpart of the entire class of the maximum likelihood estimation (MLE) solutions of the EIV model, in a 1-1 mapping. Technically, our approach gives users the flexibility to select from a class of RCR estimates the optimal one with a predefined regression efficiency criterion satisfied. Simulation studies and real-life examples are provided to illustrate the effectiveness of the RCR approach.

## 1. Introduction

Consider the linear errors-in-variables (EIV) model

$$Y = \alpha + x^T\beta + \varepsilon, \ \ X = x + \delta \tag{1}$$

where the unobservable random vector $x \in R^{p-1}$ and the linearly related random variable $y = \alpha + x^T \beta$ are subject to measurement errors, and we only observe $X$ and $Y$ instead. To ensure model identifiability, the random errors $\varepsilon$ and each component of $\delta$ are assumed to be independent and follow the same distribution. The model is called functional when $x$ is deterministic, and it is called structural when $x$ is random such that the $x$, $\varepsilon$ and $\delta$ are independent. In the present paper, we focus on the structural EIV model which is more general. A general treatment of linear EIV models is given in Fuller (1987), and more recent developments and applications are summarized in Van Huffel and Lemmerling (2002).

The popular and widely used ordinary least squares (OLS) regression is known to be biased and inconsistent in EIV models, while the orthogonal regression (OR) and geometric mean regression (GMR) are better in that case. Unfortunately, these traditional estimators can behave poorly in the presence of outliers or violations of distributional assumptions (Zamar 1989, Cheng and Vanness 1992) and therefore some robust alternatives are needed. To overcome the drawback of orthogonal regression, Brown (1982) proposed robust w-estimator which was shown to be robust to outlier contaminations in EIV models, especially in the small sample case. Ketellapper and Ronner (1984) showed robust estimation procedures, especially bounded-influence estimators, are also applicable to the EIV model with or without contaminated observational errors. Zamar (1989) proposed robust orthogonal regression M-estimators (ORM) and showed that it outperforms the robust ordinary regressions. Since the normality assumption cannot be guaranteed in real-life problems, He and Liang (2000) provided a quantile regression (QR) approach that is robust to heavier-tailed errors distribution than the Gaussian. More recently, Fekri and Ruiz-Gazen (2004 and 2006) derived a class of bounded influence robust estimators of the parameters of the EIV model from reweighted multivariate estimators of location and scatter, and extended the proposed estimators to usual error variances assumptions for the simple EIV model.

In this paper, we present a new class of robust estimators, called the robust compound regression (RCR) estimators, for the linear EIV model. This class of estimators naturally extends the least sine squares (LSS) estimator systematically studied by Han 2011. Unlike the robust regression techniques discussed in the previous paragraph and other regression M-estimators (He et al. 1990), the presented estimators not only allow for errors in both variables and departures from normality but also are robust to outliers in either response or explanatory variables. The rest of the paper is organized as follows. In Section 2, we first briefly introduce the novel LSS estimator from a motivational point of view, which serves as a prototype of the RCR-estimators. Subsequently, we define the RCR-estimators and show some of its unique properties. In Section 3, by calibrating our RCR-estimators, we compare the new robust estimators with the usual nonrobust estimators on simulated data sets as well as real-life examples. Finally, conclusions are drawn in Section 4 and some proofs are provided in Appendix.

## 2. Methodology

### 2.1 The Least Sine Squares Estimator

The RCR-estimators is actually a generalization of the LSS estimator we firstly introduce here. The LSS estimator after its name is defined to minimize the sum of squared *sine* residuals as follows

$$SS_{\text{LSS}} = \sum_{i=1}^{n} sin^2\theta_i = \sum_{i=1}^{n} \frac{d_{OR_i}^2}{d_i^2} \tag{2}$$

where $\theta_i$ denotes the angle formed by the fitted hyper-plane and the line connecting each case $(X_i^T, Y_i)$ to the dataset mean $(\bar{X}^T, \bar{Y})$. Here $d_{OR_i} = \frac{|Y_i - X_i^T\beta - \alpha|}{\sqrt{1 + \beta^T\beta}}$ represents the orthogonal residual of the $i$th observation, and the Euclidean distance from the observation to the mean is $d_i = \sqrt{(X_i - \bar{X})^T(X_i - \bar{X}) + (Y_i - \bar{Y})^2}$. Geometrically, the weighted orthogonal residual as of $d_{OR_i}/d_i$ is exactly the *sine* residual as of $sin\theta_i$.

It can be seen that the LSS estimator is simply defined through a weighted total least-squares criterion. But different from the traditional total least-squares (i.e. OR) estimator, the LSS estimator as its robust counterpart not only accounts for measurement errors on both dependent and independent variables, but also down-weighs observations with large orthogonal residuals. More importantly, from the insight of a close relationship between the total least-squares regression and the principle component analysis (PCA) approach (Jackson and Dunlevy 1988), we have the following proposition.

**Proposition 1.** Consider a data set with dependent variable $Y$ and independent variables $X = (X_{\cdot 1}, \cdots, X_{\cdot P})^T$, the LSS estimator is equivalent to the principle component associated to the smallest eigenvalue of the robust weighted sample covariance matrix:

$$\tilde{S} = \begin{bmatrix} \tilde{S}_{X_{\cdot 1}X_{\cdot 1}} & \cdots & \tilde{S}_{X_{\cdot 1}X_{\cdot P}} & \tilde{S}_{X_{\cdot 1}Y} \\ \vdots & \ddots & \vdots & \vdots \\ \tilde{S}_{X_{\cdot P}X_{\cdot 1}} & \cdots & \tilde{S}_{X_{\cdot P}X_{\cdot P}} & \tilde{S}_{X_{\cdot P}Y} \\ \tilde{S}_{YX_{\cdot 1}} & \cdots & \tilde{S}_{YX_{\cdot P}} & \tilde{S}_{YY} \end{bmatrix}_{(P+1) \text{ x } (P+1)} \tag{3}$$

where $\tilde{S}_{X_{\cdot j}X_{\cdot k}} = \sum_{i=1}^{n} \frac{(X_{ij}-\overline{X_{\cdot j}})(X_{ik}-\overline{X_{\cdot k}})}{d_i^2}$ for $j$, $k = 1, \ldots, P$, and $\tilde{S}_{X_{\cdot p}Y} = \sum_{i=1}^{n} \frac{(X_{ip}-\overline{X_{\cdot p}})(Y_i-\bar{Y})}{d_i^2}$

for $p = 1, \ldots, P$. Consequently, the LSS estimator in simple regression case yields

$$\hat{\beta}_{\text{LSS}} = \frac{\tilde{S}_{YY}-\tilde{S}_{XX}+\sqrt{(\tilde{S}_{YY}-\tilde{S}_{XX})^2+4\tilde{S}_{XY}^2}}{2\tilde{S}_{XY}}, \quad \hat{\alpha}_{\text{LSS}} = \bar{Y} - \hat{\beta}_{\text{LSS}}\bar{X} \qquad (4)$$

The proof of Proposition 1 is given in the Appendix. Of note, a closed-form solution of the LSS estimator in multivariate case can be easily obtained by the use of singular value decomposition (SVD), see Golub and Van Loan (1996).

## 2.2 The Robust Compound Regression Estimator

The RCR-estimators, as a natural extension of the LSS estimator, is defined through a weighted compound least-squares criterion by minimizing

$$SS_{\text{RCR}} = \sum_{i=1}^{n} \frac{1}{d_i^2} \left[ \gamma_0 (Y_i - \hat{Y}_i)^2 + \sum_{p=1}^{P} \gamma_p (X_{ip} - \hat{X}_{ip})^2 \right] \qquad (5)$$

subject to the constraints of $\sum_{p=0}^{P} \gamma_p = 1$ and $\gamma_p \geq 0$. Moreover, the generalized RCR can be easily defined by replacing $d_i^2$ with $d_i^{\text{k}}$ in (5), where the power $k$ can be any nonnegative integer. When $k = 0$ it becomes the objective function of ordinary compound regression, and further investigations are needed to seek an adequate power $k$ for the robustness-efficiency tradeoff. It is worth to mentioning that, in our previous work (Han 2011, Han et al. 2012), we have shown that the RCR-estimator provides a robust counterpart of the entire class of the MLE solutions of the EIV model, in a 1-1 mapping.

### 2.2.1 Parameter estimate and its asymptotic variance

Since the objective function (5) can be further simplified as

$$SS_{\text{RCR}} = \gamma_0 \sum_{i=1}^{n} \frac{(Y_i-\hat{Y}_i)^2}{d_i^2} + \sum_{p=1}^{P} \frac{\gamma_p}{\beta_p^2} \sum_{i=1}^{n} \frac{(Y_i-\hat{Y}_i)^2}{d_i^2}$$

$$= \left( \gamma_0 + \sum_{p=1}^{P} \frac{\gamma_p}{\beta_p^2} \right) \sum_{i=1}^{n} \frac{\left[ Y_i - \bar{Y} - \sum_{p=1}^{P} \beta_p (X_{ip}-\overline{X_{\cdot p}}) \right]^2}{d_i^2}$$

$$= \left( \gamma_0 + \sum_{p=1}^{P} \frac{\gamma_p}{\beta_p^2} \right) \left( \tilde{S}_{YY} + \sum_{j=1}^{P} \sum_{k=1}^{P} \beta_j \beta_k \tilde{S}_{X_{\cdot j}X_{\cdot k}} - 2 \sum_{p=1}^{P} \beta_p \tilde{S}_{X_{\cdot p}Y} \right) \qquad (6)$$

we can obtain the slope vector estimate by solving the system of $P$ equations $\frac{\partial SS_{\text{RCR}}}{\partial \beta_p} = 0$, for $p = 1, 2, \ldots, P$ simultaneously. Specifically, for any $l$ in $p = 1, \ldots, P$, we have

$$\frac{\partial SS_{\text{RCR}}}{\partial \beta_l} = -\frac{2\gamma_l}{\beta_l^3} \left( \tilde{S}_{YY} + \sum_{j=1}^{P} \sum_{k=1}^{P} \beta_j \beta_k \tilde{S}_{X_{\cdot j}X_{\cdot k}} - 2 \sum_{p=1}^{P} \beta_p \tilde{S}_{X_{\cdot p}Y} \right) +$$

$$\left( \gamma_0 + \sum_{p=1}^{P} \frac{\gamma_p}{\beta_p^2} \right) \left( 2 \sum_{\substack{p=1 \\ p \neq l}}^{P} \beta_p \tilde{S}_{X_{\cdot p}X_{\cdot l}} + 2\beta_l \tilde{S}_{X_{\cdot l}X_{\cdot l}} - 2\tilde{S}_{X_{\cdot l}Y} \right) = 0 \qquad (7)$$

Solutions can be obtained via any standard numerical software such as MATLAB, from which we have the slope estimate $\hat{\beta}_{\mathrm{RCR}}$ with some regulatory conditions satisfied. Meanwhile, the intercept estimate can be readily calculated by $\hat{\alpha}_{\mathrm{RCR}} = \bar{Y} - \bar{X}^T \hat{\beta}_{\mathrm{RCR}}$.

Obtaining asymptotic theory for the RCR-estimators in linear EIV models is a very challenging task. Hence, we use a bootstrap estimator for the asymptotic covariance matrix of the slope estimate $\hat{\beta}_{\mathrm{RCR}}$. The bootstrap estimator can be written by

$$(B-1) \sum_{b=1}^{B} (\hat{\beta}_{\mathrm{RCR}_b} - \bar{\hat{\beta}}_{\mathrm{RCR}})(\hat{\beta}_{\mathrm{RCR}_b} - \bar{\hat{\beta}}_{\mathrm{RCR}})^T \tag{8}$$

where $\hat{\beta}_{\mathrm{RCR}_b}$ is estimated from the $b$th bootstrapping sample and $\bar{\hat{\beta}}_{\mathrm{RCR}} = \frac{1}{B} \sum_{b=1}^{B} \hat{\beta}_{\mathrm{RCR}_b}$. The asymptotic variance of $\hat{\alpha}_{\mathrm{RCR}}$ can be calculated in a similar manner.

### 2.2.2 Two special members of the RCR-estimators

In the simple linear regression situation, the objective function (5) can be simplified as

$$SS_{\mathrm{RCR}} = \gamma \sum_i \frac{[Y_i - \alpha - \beta X_i]^2}{d_i^2} + (1-\gamma) \sum_i \frac{(X_i - \frac{Y_i - \alpha}{\beta})^2}{d_i^2} \tag{9}$$

where $0 \le \gamma \le 1$. Straight-forward derivations show that the slope estimate would satisfy

$$\gamma \widetilde{S_{XX}} \beta^4 - \gamma \widetilde{S_{XY}} \beta^3 + (1-\gamma) \widetilde{S_{XY}} \beta - (1-\gamma) \widetilde{S_{YY}} = 0 \tag{10}$$

Obviously at two extremes of $\gamma$, we have $\hat{\beta}_{\gamma=0} = \frac{\widetilde{S_{YY}}}{\widetilde{S_{XY}}}$ and $\hat{\beta}_{\gamma=1} = \frac{\widetilde{S_{XY}}}{\widetilde{S_{XX}}}$ respectively.

**Proposition 2.** The compound parameter $\gamma$ is a monotonic function of the slope estimate $\hat{\beta}_\gamma$, when the range of $\hat{\beta}_\gamma$ is between $\hat{\beta}_{\gamma=0}$ and $\hat{\beta}_{\gamma=1}$.

*Proof* Equation (10) can be rewritten as

$$f(\gamma) = \frac{\gamma}{1-\gamma} = \frac{\widetilde{S_{YY}} - \hat{\beta}_\gamma \widetilde{S_{XY}}}{\hat{\beta}_\gamma \widetilde{S_{XX}} - \widetilde{S_{XY}}} \frac{1}{\hat{\beta}_\gamma^3} = \frac{\hat{\beta}_{\gamma=0} - \hat{\beta}_\gamma}{\hat{\beta}_\gamma / \hat{\beta}_{\gamma=1} - 1} \frac{1}{\hat{\beta}_\gamma^3} \tag{11}$$

Note that $f(\gamma)$ is an increasing function of $\gamma \in [0,1]$. The Cauchy-Schwarz inequality ensures that $\tilde{S}_{XY}^2 \le \tilde{S}_{XX} \tilde{S}_{YY}$, hence, if $\tilde{S}_{XY} \ge 0$ we have $\hat{\beta}_{\gamma=0} \ge \hat{\beta}_\gamma \ge \hat{\beta}_{\gamma=1} \ge 0$, and thus $f(\gamma) \downarrow$ as $\hat{\beta}_\gamma \uparrow$, i.e. $\gamma$ is a decreasing function of $\hat{\beta}_\gamma$; otherwise, if $\tilde{S}_{XY} < 0$ we have $\hat{\beta}_{\gamma=0} \le \hat{\beta}_\gamma \le \hat{\beta}_{\gamma=1} < 0$, and thus $f(\gamma) \uparrow$ as $\hat{\beta}_\gamma \uparrow$, i.e. $\gamma$ is an increasing function of $\hat{\beta}_\gamma$.

Analogous to the definition of GMR in simple linear regression, the robust geometric mean (RGM) estimator is defined to minimize $-\frac{1}{2} sign(\beta) \sum_i \frac{(X_i - \hat{X}_i)(Y_i - \hat{Y}_i)}{d_i^2}$, and we can easily obtain its corresponding slope estimate $\hat{\beta}_{RGMR} = sign(\tilde{S}_{XY}) \sqrt{\frac{\tilde{S}_{YY}}{\tilde{S}_{XX}}}$.

**Proposition 3.** In the simple linear regression case, the LSS and the RGM estimators are both special members of the RCR-estimators.

*Proof*   Based on the Cauchy-Schwarz inequality, it is easy to show that $\frac{\tilde{S}_{XY}}{\tilde{S}_{XX}} \leq \hat{\beta}_{LSS} =$

$\frac{\tilde{S}_{YY} - \tilde{S}_{XX} + \sqrt{(\tilde{S}_{YY} - \tilde{S}_{XX})^2 + 4\tilde{S}_{XY}^2}}{2\tilde{S}_{XY}} \leq \frac{\tilde{S}_{YY}}{\tilde{S}_{XY}}$ given $\tilde{S}_{XY} \geq 0$, and $\frac{\tilde{S}_{YY}}{\tilde{S}_{XY}} \leq \hat{\beta}_{LSS} \leq \frac{\tilde{S}_{XY}}{\tilde{S}_{XX}}$ given $\tilde{S}_{XY} < 0$.

Similarly, we have $\frac{\tilde{S}_{XY}}{\tilde{S}_{XX}} \leq \hat{\beta}_{RGM} = \sqrt{\frac{\tilde{S}_{YY}}{\tilde{S}_{XX}}} \leq \frac{\tilde{S}_{YY}}{\tilde{S}_{XY}}$ when $\tilde{S}_{XY} \geq 0$, and $\frac{\tilde{S}_{XY}}{\tilde{S}_{XX}} \geq \hat{\beta}_{RGM} =$

$-\sqrt{\frac{\tilde{S}_{YY}}{\tilde{S}_{XX}}} \geq \frac{\tilde{S}_{YY}}{\tilde{S}_{XY}}$ when $\tilde{S}_{XY} < 0$. Since both $\hat{\beta}_{LSS}$ and $\hat{\beta}_{RGM}$ are in between $\hat{\beta}_{\gamma=0}$ and $\hat{\beta}_{\gamma=1}$, and by Proposition 2 there exists a monotonic relationship between $\gamma$ and $\hat{\beta}_\gamma$, we can conclude that both $\gamma_{LSS}$ and $\gamma_{RGM}$ are between 0 and 1. Therefore, the RCR-estimators include the LSS and the RGM estimators as special members.

### 2.2.3 Robust regression efficiency

The robust regression efficiency with respect to each regression variable is defined as the ratio of the minimized to the observed weighted squared-residuals along the corresponding coordinate direction. Mathematically, the regression efficiency with respect to the dependent variable $Y$ and each independent variable $X_{\cdot p}$ for $p = 1, \ldots, P$ are formulated as follows.

$$e_Y = \frac{min \sum_{i=1}^{n} \frac{(Y_i - \hat{Y}_i)^2}{d_i^2}}{\sum_{i=1}^{n} \frac{(Y_i - \hat{Y}_i)^2}{d_i^2}} = \frac{SS_{RCR}(\hat{\beta}_{\gamma_0=1})}{\sum_{i=1}^{n} \frac{(Y_i - \hat{Y}_i)^2}{d_i^2}} \tag{12}$$

$$e_{X_{\cdot p}} = \frac{min \sum_{i=1}^{n} \frac{(X_{ip} - \hat{X}_{ip})^2}{d_i^2}}{\sum_{i=1}^{n} \frac{(X_{ip} - \hat{X}_{ip})^2}{d_i^2}} = \frac{SS_{RCR}(\hat{\beta}_{\gamma_p=1})}{\sum_{i=1}^{n} \frac{(X_{ip} - \hat{X}_{ip})^2}{d_i^2}} \tag{13}$$

**Proposition 4.** The robust geometric mean estimator will always yield the equal $e_Y$ and $e_X$, and furthermore the maximum sum of robust regression efficiencies $e_Y + e_X$.

The proof is given in the Appendix. We hereby point out that the definition of regression efficiency is not only designed for the proposed robust compound regression analysis, but also can be utilized to compare the performance of different regression estimators given a real-data set. There are two ways to build up the corresponding goodness-of-fit criterion. One is the Frequentist approach, where all the regression variables are considered equally important due to the lack of prior knowledge, and our suggestion is the higher the sum of regression efficiencies $e_Y + \sum_{p=1}^{P} e_{X_p}$ the better the regression estimate is. The other is the Bayesian approach. For example in simple regression, if one has the prior information of keeping the prediction accuracy of $Y$ above a certain threshold, the best estimate will maximize $e_X$ subject to $e_Y \geq c$ given $c \in [0,1]$. Symmetrically, we can reverse the order of the importance for $X$ and $Y$ and obtain the best fit for $Y$ subject to $e_X \geq c$. In addition, users have the flexibility to impose constraints on more than one variable and optimize the rest in the multivariate scenario.

# 3. Simulation and Example

## 3.1 Simulation Studies

In this section, we assess the performance of our proposals in two steps. We firstly calibrate the whole class of RCR-estimators via the regression efficiency plot diagnosis, and then we focus on the comparison of the two special members – the LSS and the RGM estimators with the traditional nonrobust estimators. Simulation studies are conducted on a simple linear EIV model $Y = 1 + y + \varepsilon$ and $X = x + \delta$ with $\xi \sim N(0, 100)$, $\delta \sim N(0, \sigma_\delta^2)$ and $\varepsilon \sim N(0, \sigma_\varepsilon^2)$. In addition, we denote $\lambda = \sigma_\varepsilon^2/\sigma_\delta^2$ as the ratio of the error variances.

### *3.1.1 Calibration of the RCR-estimators by regression efficiency plot*

The robust regression efficiencies of the class of RCR-estimates can be graphically summarized through the RCR efficiency plot as shown in Figure 1. In the following scenarios, without loss of generality, we assume the error variances $\sigma_\delta^2 = \sigma_\varepsilon^2 = 10$ to be equal that is $\lambda = 1$. The model characteristics and corresponding results are summarized as follows:

(a): There are 5% outliers in the $Y$ direction with contaminated errors $\varepsilon_c \sim N(50, \sigma_\varepsilon^2)$. As can be seen from Figure 1(a), if let both $e_Y$ and $e_X$ to be no less than 0.75, we have $\gamma \in [0.471, 0.526]$ with $\hat{\beta}$ varies from 0.98 to 1.05, where the cross point corresponds to the RGM estimate $\hat{\beta} = 1.002$. The LSS estimate $\hat{\beta} = 1.004$ with corresponding $\gamma = 0.496$ falls inside the selected $\gamma$ interval.

(b): There are 5% leverage points in the $X$ direction with contaminated errors $\delta_c \sim N(50, \sigma_\delta^2)$. From Figure 1(b), we can choose $\gamma \in [0.428, 0.561]$ to make both $e_Y$ and $e_X$ to be at least 0.775, and the cross point corresponds to the RGM estimate $\hat{\beta} = 1.012$. The LSS estimate $\hat{\beta} = 1.020$ with corresponding $\gamma = 0.481$ is still inside the selected interval.

(c): Let $\xi$ follows $U(0, 100)$ distribution, and $\varepsilon$ and $\delta$ also follow uniform distributions with mean 0 such that $\sigma_\varepsilon^2/\sigma_\eta^2 = \sigma_\delta^2/\sigma_\xi^2 = 10\%$. In Figure 1(c), the range $\gamma \in$ [0.472, 0.522] ensures both e$_Y$ and e$_X$ to be at least 0.875, and it includes the LSS estimate $\hat{\beta} = 1.009$ with corresponding $\gamma = 0.491$. The cross point is the RGM estimate $\hat{\beta} = 1.007$.

(d): Let $\xi$ follows $\sqrt{5}t(3)$ distribution, and $\varepsilon$ and $\delta$ both follow the student's $t(3)$ distribution such that $\sigma^2_\varepsilon/\sigma^2_\eta = \sigma^2_\delta/\sigma^2_\xi = 20\%$. By setting both e$_Y$ and e$_X$ above a threshold of 0.725 as shown in Figure 1(d), we find $\gamma \in [0.445, 0.529]$ and the RGM estimate $\hat{\beta} = 1.027$. Again, the LSS estimate $\hat{\beta} = 1.055$ with corresponding $\gamma = 0.447$ are within the selected $\gamma$ interval.

We further apply the bootstrap resampling technique with 1000 replicates to obtain the 95% confidence interval (C.I.) of the RCR estimate $\hat{\beta}$. As can be seen from Figure 2, the true slope $\beta = 1$ always lies in the 95% C.I. of the selected RCR slope estimates.

<div align="center">**(Insert Figure 1 and Figure 2)**</div>

### *3.1.2 Comparison of different estimators*

Since the LSS and the RGM estimators are most representative among the whole class of RCR-estimators, a more extensive simulation study is necessary to systematically compare them with their nonrobust counterparts – the OR and GMR estimators. We keep the above settings of the linear EIV model unchanged, and draw 10000 samples of size $n = 200$ from that model. Since all the fitted regression lines can be expressed by the point-slope form $Y - \bar{Y} = \hat{\beta}(X - \bar{X})$, we therefore expect that the results of simulation for the intercept are similar to those of the slope.

To measure performance, we use the bias, the standard deviation (std) as well as the root mean squared error (RMSE) of the $\hat{\beta}$. Since the ratio of the error variances $\lambda$ is known, the MLE solution of structural model is treated as the ground truth. Of note, the MLEs are calculated after excluding the outliers for the contaminated EIV data studies. The results of our simulation study are given in Tables 1-4. For the uncontaminated EIV data in the first two tables, our estimators are comparable with the traditional EIV model estimator. More importantly, the last two tables show good performance of our new estimators in the presence of outliers, while the OR and GMR estimators are badly influenced by such contaminations.

<div align="center">**(Insert Tables 1-4)**</div>

### **3.2 Real-life Examples**

### *3.2.1 Serum kanamycin data*

To further illustrate and motivate our proposals, let us consider a simple example which is given by (Kelly 1984) and reanalyzed by (Zamar 1989) in the context of EIV models. The data consists of simultaneous pairs of measurements of serum kanamycin levels in blood samples drawn from twenty babies. One of the measurements was obtained by a heelstick method ($X$), the other by using an umbilical catheter ($Y$). The question was whether the two methods are systematically different and so that one could be substituted for the other after correction for bias. It seems reasonable to assume that both methods are subject to measurement errors with equal variances (Kelly 1984). To better test the robustness of different estimator, we change the original value (33.2, 26.0) of case 2 to (39.2, 32.0) as in the numerical example given by (Zamar 1989).

Using the usual OR and the GMR estimators lead to about 0.97 for the slope while using the LSS estimator we propose leads to 1.52 for the slope, which is very similar to the slope (1.39) of Zamar's orthogonal regression M-estimators (ORM). We apply the RCR-estimators on this data set as well, and as can be seen from Figure 4(a), if we set both $e_Y$ and $e_X$ be no less than 0.825, the satisfied $\gamma$ interval is [0.344, 0.379] and the corresponding slope ranges from 1.32 to 1.34. Of note, the results by previous authors show that observations 2, 16 are potential influential, and furthermore observation 2 has much greater influence on the slope compared to case 16. Figure 3(b) depicts the residual plot from the RGM estimate with maximum regression efficiency. It is clear that observation 2 is very influential and observation 16 has little influence.

**(Insert Figure 3)**

### 3.2.2 Brain versus body weights data

Another numerical example is provided to illustrate the effectiveness of the RCR-estimators on the data set given in Rousseeuw and Leroy (1987). The data has been analyzed by He and Liang (2000) and Fekri and Ruiz-Gazen (2006). The data consists of brain and body weights of 28 animals. We also take the view that both weights are assumed to be measured with error and some observations are outlying (Fekri and Ruiz-Gazen 2006). The question we are interested in is whether a larger brain is required to govern a heavier body, or from another perspective, whether the brain weight increases linearly as the body weight increases.

We obtain the usual OR estimates $\hat{\alpha}_{OR} = 2.29, \hat{\beta}_{OR} = 0.57$ and the GMR estimates $\hat{\alpha}_{GMR} = 2.03, \hat{\beta}_{GMR} = 0.64$. Compared to both, the LSS estimates $\hat{\alpha}_{LSS} = 1.84, \hat{\beta}_{LSS} = 0.68$ are more close to the slope estimate $\hat{\beta}_{QR} = 0.74$ from He and Liang's 50% quantile regression (QR). In addition, we also apply the RCR-estimators on this data set, and as can be seen from Figure 5(a), if we set both $e_Y$ and $e_X$ to be no less than 0.75, the satisfied $\gamma$ interval is [0.560, 0.645], and the corresponding slope estimate varies from 0.79 to 0.83. Cases 6, 16 and 25 are detected by the RGM estimator in Figure 4(b), because they as dinosaurs are distinct from the other 25 cases of mammals.

**(Insert Figure 4)**

# 4. Summary and Conclusion

In the present paper, we propose a novel robust estimator for the multivariate linear EIV model. We not only show the LSS estimator is a special case of the RCR-estimators, but also prove the optimality of the RGM in the respect of maximizing the sum of regression efficiencies in simple linear regression. The generalized RCR-estimators is defined for further investigations, and the sum of regression efficiencies as a general goodness-of-fit criterion is proposed to compare the estimates from different regression approaches in real data analysis.

The advantages of the new RCR-estimators lie in its intuitive geometric representation, its distribution free nonparametric nature being a direct generalization of the nonparametric compound regression analysis method, its operational independence to the ratio of the error variances, and more importantly its robustness to outliers and other departures from the underlying assumptions. Nevertheless, the disadvantage of our proposals is that the fitted hyper-plane must pass through the center of the target dataset which will to some extent sacrifice its robustness performance. A robust location estimator from the data depth point of view (Liu et al. 1999) could be a remedy to further enhance the robustness of the newly proposed RCR-estimators.

# Appendix

## A.1. Proof of Proposition 1

Without loss of generality, we form the multivariate regression model as $\sum_{p=1}^{P} \beta_p X_p = 0$ or in matrix form $X\beta = 0$ for the centered data, where $X = [X_1, X_2, \ldots, X_P]$ is a $n$ by $P$ matrix of observations, and $\beta$ is a $P$ by 1 column vector of regression coefficients. This linear relationship is uniquely specified by imposing the constraint $\beta^{\mathrm{T}}\beta = 1$.

We first define $\bar{X} = [\bar{X}_1, \bar{X}_2, \ldots, \bar{X}_P] = \begin{bmatrix} \frac{X_{11}}{d_1} & \cdots & \frac{X_{p1}}{d_1} \\ \vdots & \ddots & \vdots \\ \frac{X_{1n}}{d_n} & \cdots & \frac{X_{Pn}}{d_n} \end{bmatrix}_{n \times P}$ as the $n$ by $P$ transformed matrix of observations, where $d_i = \sqrt{\sum_{p=1}^{P} X_{pi}^2}$ is the distance from the $i$th observation to

the origin. In this context, the LSS is defined to minimize

$$SS_{\text{LSS}} = (\bar{X}\beta)^T (\bar{X}\beta) = \beta^T \bar{X}^T \bar{X}\beta = \beta^T (\bar{X}^T \bar{X})\beta = \beta^T \bar{S}\beta$$

where $\bar{S}$ is the $P$ by $P$ robust sample covariance matrix

$$\bar{S} = \begin{bmatrix} \bar{X}_1^T \bar{X}_1 & \cdots & \bar{X}_1^T \bar{X}_P \\ \vdots & \ddots & \vdots \\ \bar{X}_P^T \bar{X}_1 & \cdots & \bar{X}_P^T \bar{X}_P \end{bmatrix}_{P \times P}$$

We define the eigenvectors of $\bar{S}$ as $(\alpha_1, \cdots, \alpha_P)$ in the order of descending eigenvalues $(\lambda_1, \cdots, \lambda_P)$. Assume $\bar{S}$ is non-singular, the eigenvectors can expand the $P$-dimensional space, and then the slope estimate $\beta$ can be expressed as a linear combination $(l_1\alpha_1 + \cdots + l_P\alpha_P)$ subject to $\sum_{p=1}^{P} l_p = 1$. Hence, the minimization of $SS_{\text{LSS}}$ is equivalent to the minimization of

$$(l_1\alpha_1 + \cdots + l_P\alpha_P)^T \bar{S}(l_1\alpha_1 + \cdots + l_P\alpha_P)$$

Since as we know that the eigenvectors are orthogonal and $\alpha_p^T \bar{S}\alpha_p = \lambda_p$, the problem becomes the minimization of $\sum_{p=1}^{P} \lambda_p l_p$. Under the constraints $\sum_{p=1}^{P} l_p = 1$ and $\lambda_1 \geq \cdots \geq \lambda_P$, the minimum is achieved when we set $l_P = 1$ and thus $\beta = \alpha_P$. That is the eigenvector corresponding to $\lambda_P$ - the smallest eigenvalue of $\bar{S}$.

## A.2. Proof of Proposition 4

(1) The RGM estimator aims to minimize

$$SS_{\text{RGM}} = -\frac{1}{2} sign(\beta) \sum_i \frac{(X_i - \hat{X}_i)(Y_i - \hat{Y}_i)}{R_i^2} = -\frac{1}{2} sign(\beta) \sum_i \frac{[(X_i - \bar{X}) - \frac{1}{\beta}(Y_i - \bar{Y})][(Y_i - \bar{Y}) - \beta(X_i - \bar{X})]}{R_i^2}$$

By solving $\frac{\partial SS_{\text{RGM}}}{\partial \beta} = 0$, we obtain the slope estimate $\hat{\beta} = sign(\bar{S}_{XY})\sqrt{\bar{S}_{YY}/\bar{S}_{XX}}$.

Given the following conditions

$$\sum_{i=1}^{n}(Y_i - \hat{Y}_i)^2 = \bar{S}_{YY} + \hat{\beta}^2 \bar{S}_{XX} - 2\hat{\beta}\bar{S}_{XY}$$

$$\sum_{i=1}^{n}(X_i - \hat{X}_i)^2 = \frac{1}{\hat{\beta}^2}\sum_{i=1}^{n}(Y_i - \hat{Y}_i)^2 = \frac{1}{\hat{\beta}^2}\bar{S}_{YY} + \bar{S}_{XX} - \frac{2}{\hat{\beta}}\bar{S}_{XY} ,$$

the regression efficiencies $e_Y$ and $e_X$ can be further expressed as

$$e_Y = \frac{min\sum_{i=1}^{n}(Y_i - \hat{Y}_i)^2}{\sum_{i=1}^{n}(Y_i - \hat{Y}_i)^2}$$

$$= \frac{\sum_{i=1}^n (Y_i - \hat{Y}_i)^2|_{\hat{\beta} = \tilde{S}_{XY}/\tilde{S}_{XX}}}{\sum_{i=1}^n (Y_i - \hat{Y}_i)^2|_{\hat{\beta} = sign(\tilde{S}_{XY})\sqrt{\tilde{S}_{YY}/\tilde{S}_{XX}}}}$$

$$= \frac{\tilde{S}_{XX}\tilde{S}_{YY} - \tilde{S}_{XY}}{2\tilde{S}_{XX}\tilde{S}_{YY} - 2|\tilde{S}_{XY}|\sqrt{\tilde{S}_{XX}\tilde{S}_{YY}}}$$

and

$$e_X = \frac{min \sum_{i=1}^n (X_i - \hat{X}_i)^2}{\sum_{i=1}^n (X_i - \bar{X}_i)^2}$$

$$= \frac{\frac{1}{\hat{\beta}^2} \sum_{i=1}^n (Y_i - \hat{Y}_i)^2|_{\hat{\beta} = \tilde{S}_{YY}/\tilde{S}_{XY}}}{\frac{1}{\hat{\beta}^2} \sum_{i=1}^n (Y_i - \hat{Y}_i)^2|_{\hat{\beta} = sign(\tilde{S}_{XY})\sqrt{\tilde{S}_{YY}/\tilde{S}_{XX}}}}$$

$$= \frac{\tilde{S}_{XX}\tilde{S}_{YY} - \tilde{S}_{XY}}{2\tilde{S}_{XX}\tilde{S}_{YY} - 2|\tilde{S}_{XY}|\sqrt{\tilde{S}_{XX}\tilde{S}_{YY}}} \,.$$

Hence, we have shown the equality of regression efficiencies $e_Y$ and $e_X$.

(2) For the slope estimate $\hat{\beta}$ of any estimator, we have

$$E = e_Y + e_X$$

$$= \frac{\sum_{i=1}^n (Y_i - \hat{Y}_i)^2|_{\hat{\beta} = \tilde{S}_{XY}/\tilde{S}_{XX}}}{\sum_{i=1}^n (Y_i - \hat{Y}_i)^2} + \frac{\frac{1}{\hat{\beta}^2} \sum_{i=1}^n (Y_i - \hat{Y}_i)^2|_{\hat{\beta} = \tilde{S}_{YY}/\tilde{S}_{XY}}}{\frac{1}{\hat{\beta}^2} \sum_{i=1}^n (Y_i - \hat{Y}_i)^2}$$

$$= \frac{(\tilde{S}_{YY} - \tilde{S}_{XY}/\tilde{S}_{XX}) + \beta^2(\tilde{S}_{XX} - \tilde{S}_{XY}/\tilde{S}_{YY})}{\tilde{S}_{YY} + \beta^2 \tilde{S}_{XX} - 2\beta \tilde{S}_{XY}} \,.$$

Let $\frac{\partial \Sigma}{\partial \hat{\beta}} = 0$, we have $\hat{\beta}^2 = \frac{\tilde{S}_{YY}}{\tilde{S}_{XX}}$, which means $\Sigma = e_Y + e_X$ is unimodal and maximized at

$\hat{\beta} = \sqrt{\frac{\tilde{S}_{YY}}{\tilde{S}_{XX}}} = sign(\tilde{S}_{XY})\sqrt{\frac{\tilde{S}_{YY}}{\tilde{S}_{XX}}} = \hat{\beta}_{\text{RGM}}$ when $\tilde{S}_{XY} \geq 0$, and it is unimodal and maximized at $\hat{\beta} = -\sqrt{\frac{\tilde{S}_{YY}}{\tilde{S}_{XX}}} = sign(\tilde{S}_{XY})\sqrt{\frac{\tilde{S}_{YY}}{\tilde{S}_{XX}}} = \hat{\beta}_{\text{RGM}}$ when $\tilde{S}_{XY} < 0$.

Therefore, we have proven that the RGM estimator has the maximum $e_Y + e_X$.

RCR: Robust Compound Regression

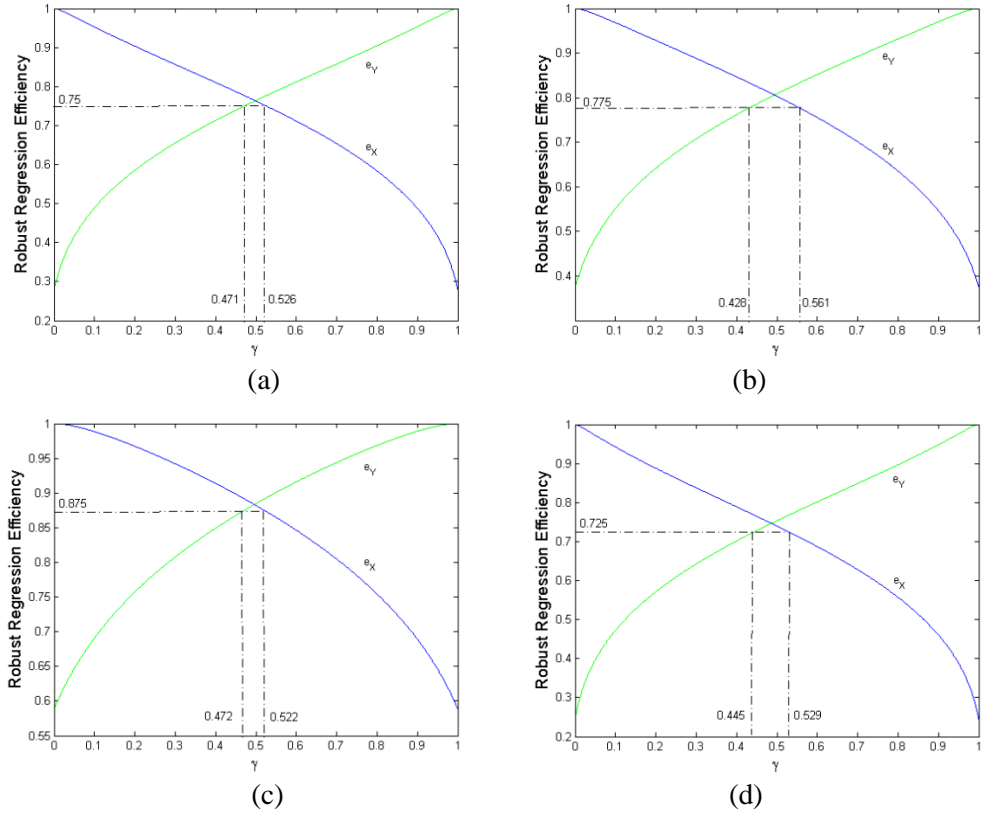

**Figure 1.** The RCR efficiency plots of (a) EIV data with 5% outliers, (b) with 5% leverage points, (c) with uniformly distributed errors, and (d) with student's t distributed errors.

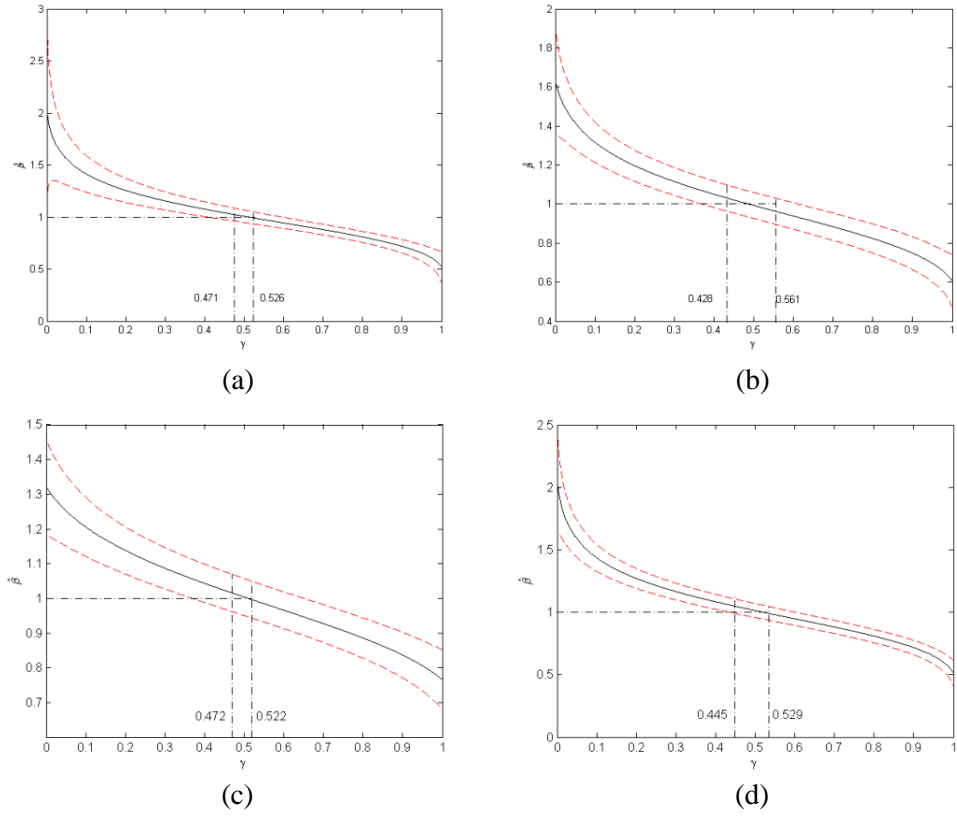

**Figure 2.** The 95% C.I. of RCR slope estimates of (a) EIV data with 5% outliers, (b) with 5% leverage points, (c) with uniformly distributed errors, and (d) with student's t distributed errors.

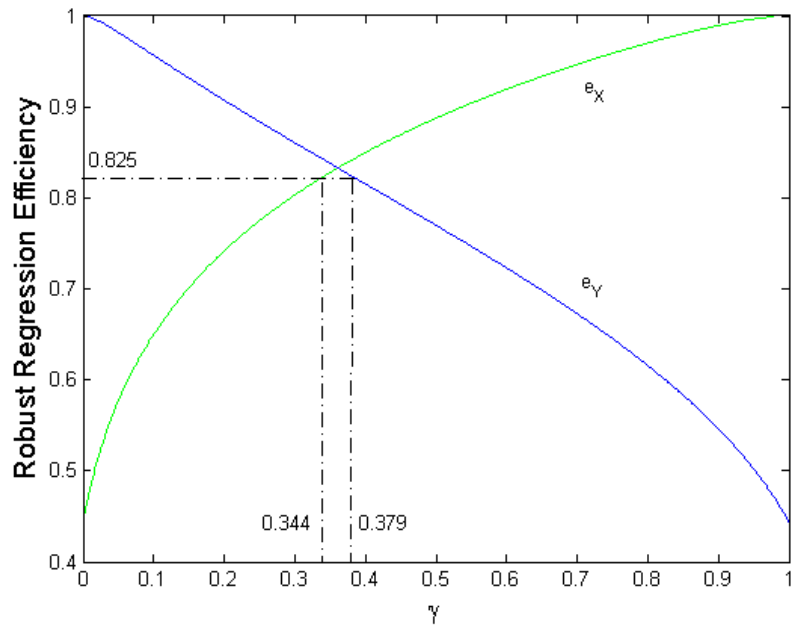

(a)

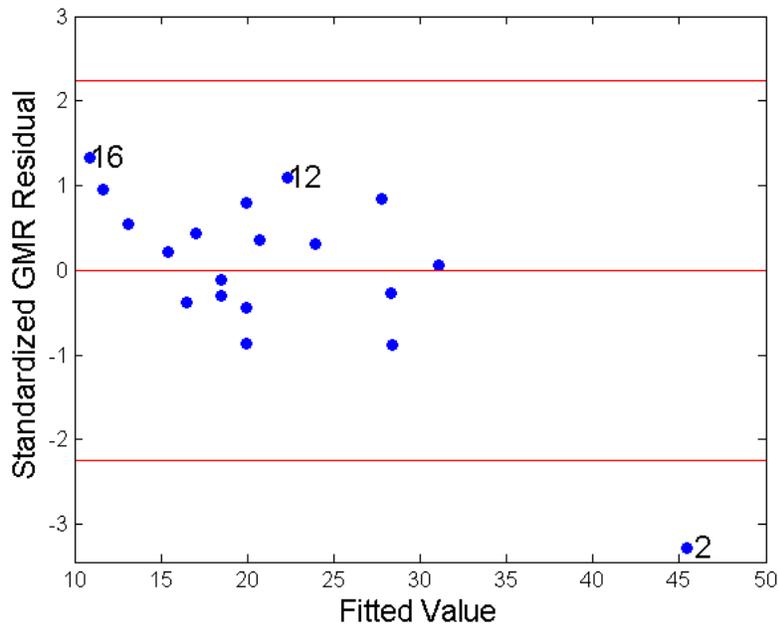

(b)

**Figure 3.** (a) The RCR efficiency plot and (b) the RGM residual diagnosis plot on the serum kanamycin data set

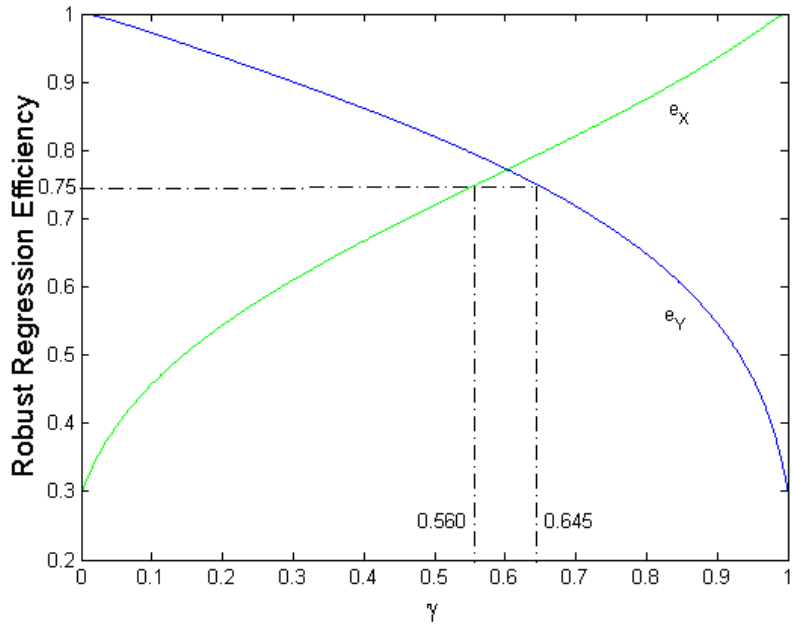

(a)

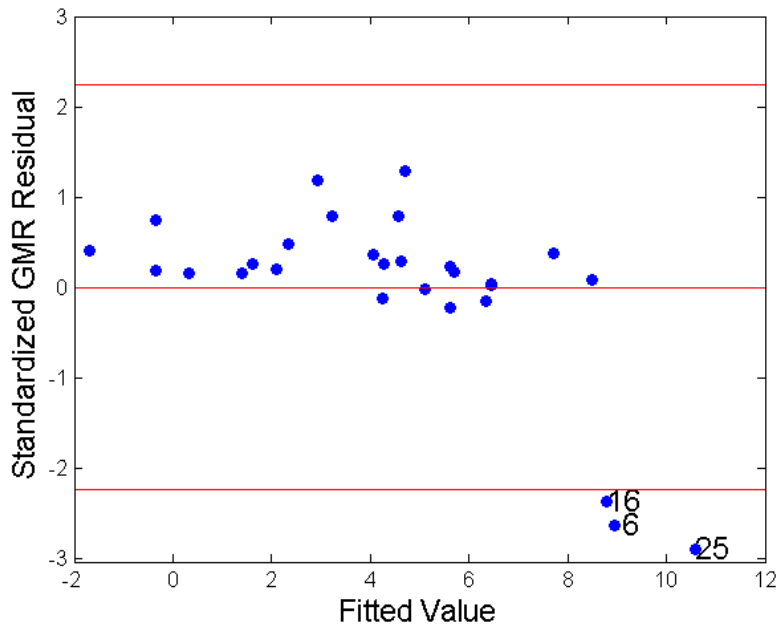

(b)

**Figure 4.** (a) The RCR efficiency plot and (b) the RGM residual diagnosis plot on the brain versus body weights data set

**Table 1:** Comparison of MSE, bias, std from different estimators on uncontaminated EIV data with different $\lambda$ and the *X* direction noise-to-signal ratio is .2

| $\lambda$ | $RMSE(\hat{\beta})$ | | | | $bias(\hat{\beta})$ | | | | $std(\hat{\beta})$ | | | |
|---|---|---|---|---|---|---|---|---|---|---|---|---|
| | OR | LSS | GMR | RGM | OR | LSS | GMR | RGM | OR | LSS | GMR | RGM |
| 0 | .097 | .106 | .089 | .072 | -.094 | -.093 | -.087 | -.062 | .029 | .054 | .026 | .036 |
| .25 | .074 | .088 | .066 | .058 | -.071 | -.070 | -.064 | -.044 | .034 | .060 | .031 | .038 |
| .5 | .050 | .072 | .044 | .047 | -.048 | -.046 | -.042 | -.028 | .038 | .066 | .033 | .039 |
| .75 | .025 | .064 | .023 | .041 | -.024 | -.023 | -.021 | -.014 | .043 | .074 | .036 | .041 |
| 1 | 0 | .068 | .008 | .040 | .001 | .003 | .001 | .001 | .048 | .080 | .040 | .043 |

**Table 2:** Comparison of MSE, bias, std from different estimators on uncontaminated EIV data with different $\lambda$ and the *X* direction noise-to-signal ratio is .05

| $\lambda$ | $RMSE(\hat{\beta})$ | | | | $bias(\hat{\beta})$ | | | | $std(\hat{\beta})$ | | | |
|---|---|---|---|---|---|---|---|---|---|---|---|---|
| | OR | LSS | GMR | RGM | OR | LSS | GMR | RGM | OR | LSS | GMR | RGM |
| 1 | 0 | .042 | .001 | .031 | .000 | .001 | .000 | .001 | .023 | .047 | .022 | .034 |
| 2 | .026 | .055 | .024 | .038 | .026 | .028 | .024 | .018 | .029 | .056 | .027 | .038 |
| 3 | .052 | .076 | .047 | .049 | .051 | .053 | .047 | .034 | .034 | .062 | .031 | .039 |
| 4 | .079 | .100 | .070 | .061 | .078 | .081 | .070 | .048 | .039 | .071 | .035 | .042 |
| 5 | .107 | .126 | .093 | .072 | .107 | .108 | .092 | .061 | .044 | .077 | .038 | .043 |

**Table 3:** Comparison of MSE, bias, std from different estimators on contaminated EIV data with different $\lambda$ and 5% outliers in the *X* direction

| $\lambda$ | $RMSE(\hat{\beta})$ | | | | $bias(\hat{\beta})$ | | | | $std(\hat{\beta})$ | | | |
|---|---|---|---|---|---|---|---|---|---|---|---|---|
| | OR | LSS | GMR | RGM | OR | LSS | GMR | RGM | OR | LSS | GMR | RGM |
| 0 | .459 | .149 | .343 | .098 | -.458 | -.142 | -.341 | -.093 | .036 | .048 | .032 | .032 |
| .5 | .445 | .133 | .324 | .083 | -.444 | -.125 | -.323 | -.077 | .036 | .054 | .032 | .034 |
| 1 | .436 | .118 | .310 | .071 | -.434 | -.107 | -.309 | -.063 | .040 | .061 | .034 | .036 |
| 1.5 | .422 | .101 | .294 | .058 | -.421 | -.085 | -.293 | -.048 | .043 | .067 | .035 | .037 |
| 2 | .410 | .088 | .280 | .052 | -.406 | -.062 | -.275 | -.033 | .047 | .073 | .036 | .039 |

**Table 4:** Comparison of MSE, bias, std from different estimators on contaminated EIV data with different $\lambda$ and 5% outliers in the *Y* direction

| $\lambda$ | $RMSE(\hat{\beta})$ | | | | $bias(\hat{\beta})$ | | | | $std(\hat{\beta})$ | | | |
|---|---|---|---|---|---|---|---|---|---|---|---|---|
| | OR | LSS | GMR | RGM | OR | LSS | GMR | RGM | OR | LSS | GMR | RGM |
| 0 | .704 | .075 | .420 | .047 | .695 | .053 | .415 | .032 | .115 | .056 | .067 | .034 |
| .5 | .740 | .106 | .436 | .062 | .732 | .088 | .431 | .051 | .115 | .067 | .067 | .039 |
| 1 | .778 | .133 | .452 | .075 | .769 | .116 | .447 | .064 | .126 | .072 | .068 | .039 |
| 1.5 | .823 | .165 | .467 | .087 | .814 | .151 | .464 | .079 | .136 | .082 | .073 | .042 |
| 2 | .864 | .201 | .483 | .101 | .854 | .187 | .479 | .093 | .146 | .089 | .075 | .043 |